\documentclass[3p]{elsarticle}
\makeatletter
\def\ps@pprintTitle{%
 \let\@oddhead\@empty
 \let\@evenhead\@empty
 \def\@oddfoot{\centerline{\thepage}}%
 \let\@evenfoot\@oddfoot}
\makeatother

\usepackage{amssymb}

\usepackage{graphicx}
\usepackage{balance}
\usepackage{color}
\usepackage{xcolor}
\usepackage{multirow}
\usepackage{caption}
\usepackage{enumitem}
\usepackage{array}

\usepackage{subfigure,epsfig,amsfonts,amsmath,cases,amssymb,graphics, latexsym, times, algorithmic, algorithm, bbm, color, soul,wrapfig,comment,longtable}

\begin{document}

\begin{frontmatter}



\title{Personal Mental Health Navigator: Harnessing the Power of Data, Personal Models, and Health Cybernetics to Promote Psychological Well-being}



\author[uci-CS,uci-NS]{Amir M. Rahmani\corref{cor1}}
\ead{a.rahmani@uci.edu}

\author[uci-PS]{Jocelyn Lai}

\author[uci-CS]{Salar Jafarlou}

\author[uci-PS]{Asal Yunusova}

\author[uci-PS]{Alex. P. Rivera}

\author[uci-CS]{Sina Labbaf}

\author[uci-S,uci-E]{Sirui Hu}

\author[turku]{Arman Anzanpour}

\author[uci-CS]{Nikil Dutt}

\author[uci-CS]{Ramesh Jain}

\author[uci-PS]{Jessica L. Borelli}

\address[uci-CS]{Department of Computer Science, University of California, Irvine, USA}
\address[uci-NS]{School of Nursing, University of California, Irvine, USA}
\address[uci-PS]{Department of Psychological Science, University of California, Irvine, USA}
\address[uci-S]{Department of Statistics, University of California, Irvine, USA}
\address[uci-E]{Department of Economics, University of California, Irvine, USA}
\address[turku]{Department of Future Technologies, University of Turku, Turku, Finland}
\cortext[cor1]{Corresponding author}

\begin{abstract}
Traditionally, the regime of mental healthcare has followed an episodic psychotherapy model wherein patients seek care from a provider through a prescribed treatment plan developed over multiple provider visits.  
Recent advances in wearable and mobile technology have generated increased interest in digital mental healthcare that enables individuals to address episodic mental health symptoms such as depression and anxiety.  
However, despite providers' best intentions, these efforts are typically reactive and symptom-focused, and do not provide comprehensive, wrap-around, customized treatments that capture an individual’s holistic mental health model as it unfolds over time.
Recognizing that each individual is unique and requires personally tailored mental health treatment, we present the notion of Personalized Mental Health Navigation (MHN): a therapist-in-the-loop, cybernetic goal-based system that deploys a continuous cyclic loop of measurement, estimation, guidance, to steer the individual’s mental health state towards a healthy zone. 
We outline  the major components of MHN that is premised on the development of an individual’s  personal mental health state, holistically represented by a high-dimensional cover of multiple knowledge layers such as emotion, biological patterns, sociology, behavior, and cognition. 
We demonstrate the feasibility of the personalized MHN approach via a 12-month pilot case study for holistic stress reduction and management in college students, and highlight an instance of a therapist-in-the-loop intervention using MHN for monitoring, estimating, and proactively addressing moderately severe depression over a sustained period of time.
We believe MHN paves the way to  transform mental healthcare from the current passive, episodic, reactive process (where individuals seek help to address symptoms that have already manifested) to a continuous and navigational paradigm that leverages a personalized model of the individual, promising to deliver timely  interventions to individuals in a holistic manner.

\end{abstract}


\end{frontmatter}

\vspace{-6pt}
\section{Introduction} \label{introduction}

Mental health is an important factor in determining an individual's quality of life. While it can directly affect the quality of life, mental health can also have indirect effects, for instance by changing the ways in which individuals engage in decision making processes, resulting in potential long-term effects across the lifespan \cite{bishop2018anxiety,hockey2000effects}. 
Recognizing that each person is unique, the recent P4  medicine approach \cite{P4} aims to transform the practice of medicine from a  traditionally reactive, symptomatic  approach to a proactive systems approach that addresses the causes via predictive, preventive, personalized and participatory strategies.
The current mental healthcare system similarly deploys an acute and symptom-focused reactive approach to patient well-being. 
Healthcare providers often intervene after symptoms have already manifested within an individual, as opposed to adopting an approach that seeks to prevent illness from developing or sustain well-being. One major drawback of this system is its passive approach to mental health. Indeed, in many cases individuals only become conscious of their issues once their conditions become severe or reach a point where they perceive a need for the issue to be addressed \cite{elhai2009sociodemographic,lindsay2009role}. In this passive, reactive model, there would be little effort to monitor one's own behavior or experiences as well as actively seek guidance in the absence of conscious discomfort. 

Furthermore, the traditional episodic medical and psychotherapy model of treatment for mental health is premised on the notion that providers interact with the individual during scheduled appointments potentially few and far between, or otherwise for prearranged circumscribed amounts of time per week. The provider relies upon the individual to be an accurate reporter of their symptoms and health, both in the present moment and over a period of time (i.e., between appointments or sessions), and to summarize their experiences using language the provider will understand with minimal misinterpretations (i.e., shared frame of reference for tolerable levels of pain, typical levels of activity, what "triggers" means, etc.). Through this brief time frame, providers must synthesize what information they receive from the individual in order to prescribe a treatment plan. Thus, the provider is expected to reference information provided during the session through interaction and observations to find the causes and provide recommendations to alleviate them. This model of mental health treatment places a large burden of responsibility on the provider (to synthesize a large amount of information in short span of time) as well as on the individual (to accurately and honestly appraise and report their experiences to the provider). There are considerable limitations to this approach as both provider and individual are limited in their ability to fulfill these obligations.

An additional limitation to the current mental health system is the lack of sufficient holistic information about individuals in their daily life in terms of the nuanced fluctuations in mental well-being across days, their daily activities, and the context in which they experience varying emotions and symptoms \cite{carter2007momentary,myin2009experience}. The primary source of information exchanged between a healthcare provider and the individual is often the individual's subjective experience and memories. Although having individuals describe their daily schedules can help reveal important patterns that may not otherwise be shared, there may be the possibility that the individual forgets or over-generalizes their daily routine, or may choose to hide aspects of their experiences they are uncomfortable disclosing. In recent years, the field of mental health has increasingly realized the inherent flaws in this methodology in the sense that psychological states are not only subjective and culturally constrained  \cite{betancourt2005cultural,koehn2006medical,schouten2006cultural}, but that there are also reasons to be concerned about this method of reporting, such as the fact that people have poor memory regarding their behavior and experiences \cite{schmier2004patient,stull2009optimal}, poor insight into the cause of their behavior \cite{vazire2008knowing}, and when asked to provide explanations for behavior they do not understand, will engage in confabulation without being aware of it \cite{coltheart2017confabulation}. Understanding the patient's experience may also be a difficult feat for the provider due to challenges to the patient-provider relationship (i.e., trust, cultural competency). 
The lack of holistic and comprehensive information about the individual's experience may ultimately result in less-optimal care delivered.

Recent shifts in health models and improvements in technology have offered complementary approaches to addressing these limitations, by incorporating an individual's unique genetics, lifestyle and exposome to offer personalized interventions.
These trends are championed by the P4 \cite{P4} (predictive, preventive, personalized, and participatory) paradigm that aims to look at health and medicine based on all knowledge and data that has now become commonly available. 
Indeed, the 
field of medicine 
has long adapted the biopsychosocial model as a holistic approach towards understanding health and care. 
The biopsychosocial model takes into account both external (i.e., sociocultural) and internal factors (i.e., biology and psychology) that may impact health \cite{borrell2004biopsychosocial,engel1981clinical,wade2017biopsychosocial}. 
External factors encompass a wide-range of contexts including but not limited to family, community environment, and larger societal structures. Dynamic interactions across these numerous factors may relate to the ways in which one's health fluctuates across a continuum of well-being and quality of life. 
Keyes \cite{keyes2002mental} suggested a mental health continuum where individuals may vary both objectively and subjectively in their state of well-being, where one end of the continuum reflects higher quality of life and flourishing, and the opposing end reflects worsened quality of life and presence of mental health issues.
Indeed, recent trends have included research and interventions aimed at sustaining flourishing and well-being as a preventative approach to mental health. 
With advances in ubiquitous and wearable sensing, a change in an individual's position on the mental health continuum can be tracked, offering tremendous opportunities and greater insight to mental health professionals to better understand an individual's needs and navigate them towards a healthy state of mind through personalized monitoring and intervention. Depending on the severity and type of the intervention, such services can be provided through a therapist-in-a-loop model or autonomously using smart guidance systems (see Figure 1). 

\begin{figure}[h]
  \centering
  \includegraphics[width=0.85\linewidth]{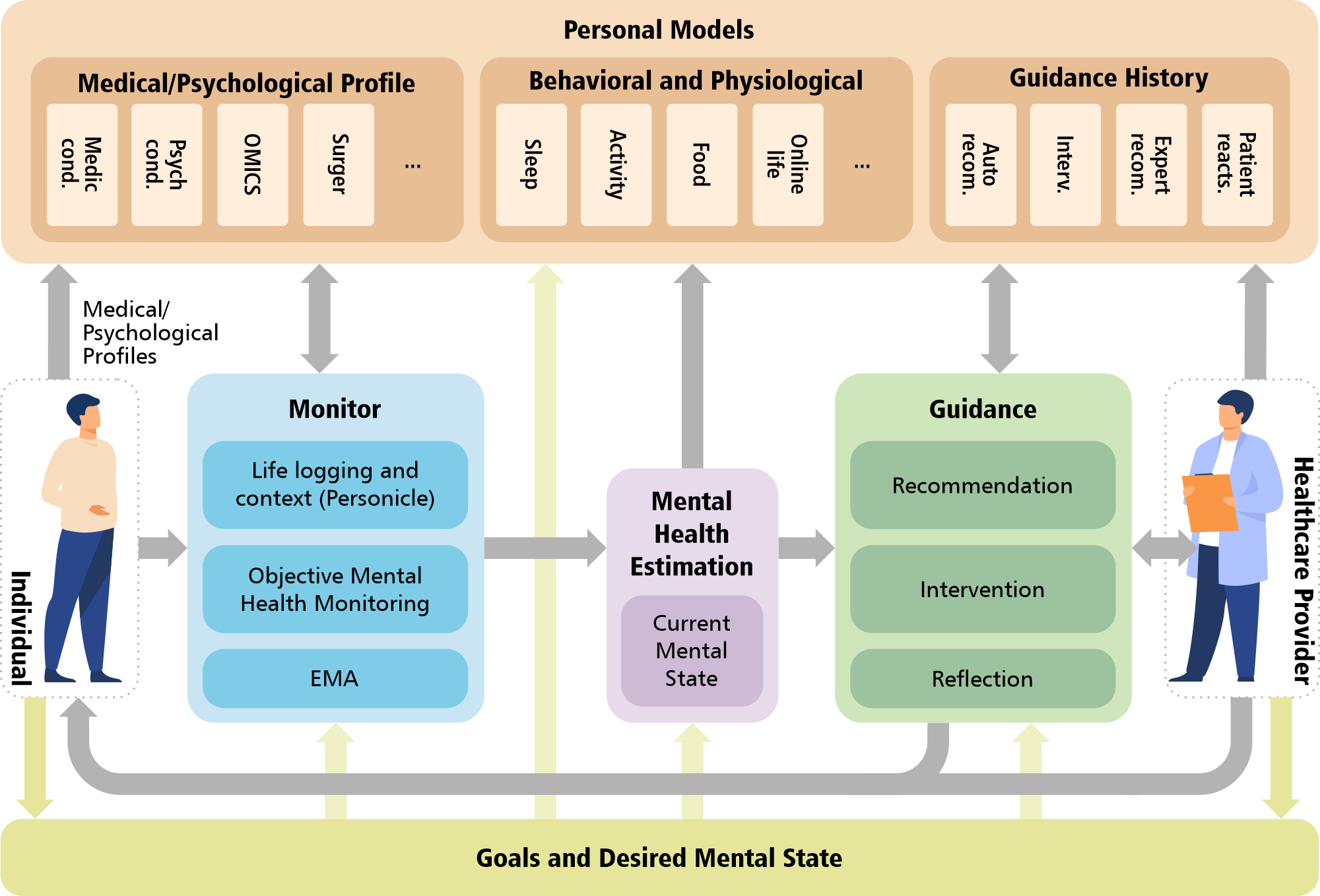}
  \caption{A Personalized Mental Health Navigator System Using a Continuous Cyclic Loop} 
  \vspace{-5mm}
  \label{fig:mhn}
\end{figure}

In this paper, we propose the notion of Personalized Mental Health Navigation (MHN) as a goal-based closed-loop guidance system that has the potential to transform mental health 
care from reactive and  episodic
to a continuous and navigational paradigm that leverages a holistic, personalized model of the individual. To demonstrate the versatility of this approach, we also present a case study on applying the personalized MHN approach for holistic stress reduction and management in college students in the service of impacting their mental health. As a group, college students are among the most voracious consumers of technology, in part because they grew up using smart phones, with some data even suggesting teenagers text more than they speak \cite{garcia2014,nielsen2010}. As such, they are ideal targets for the investigation of scientific questions related to life tracking, as well as targets of interventions aimed at reducing stress through technology.


\section{The Personalized Mental Health Navigator Model}

The Personalized Mental Health Navigator Model can best be illustrated with a simple metaphor: imagine using a route navigator software (e.g., Google Maps) to navigate you from your current location to your destination. The navigator software constantly monitors your location using GPS, and estimates your current state based on your current latitude and longitude on the map.
The navigator then identifies the most efficient route (based on your preference) and gives you step-by-step guidance on how to reach your destination. If you make a wrong turn or decide to make a stop, the navigator will reflect on your decision almost immediately and correct your path. Route navigator is an operational example of  a cybernetic feedback control system \cite{complex}. 

Inspired by the route navigator model, we propose a personalized MHN system with similar components in the mental health domain. Figure \ref{fig:mhn} demonstrates a flow diagram of the proposed MHN system including five major components integrated in a cybernetics 
\cite{Cybernetics}
structure: \textit{i)} Goals (i.e., the desired mental state), \textit{ii)} Monitor, \textit{iii)} Mental Health Estimation, \textit{iv)} Personal Models, and \textit{v)} Guidance. The cybernetics approach allows for a continuous cyclic loop of measurement, estimation, guidance, and influence to monitor and make sure that the person’s mental health state remains in a healthy zone. 

The science of cybernetics is centered around setting \textit{Goals} and devising action sequences to accomplish and maintain those goals in the presence of noise and disturbances. 
In the context of mental health, an individual's life challenges can be considered as disturbances. 
In mental health navigation, goal setting is not a fixed state, but rather a collaborative process between an individual and a healthcare provider to reach a goal consensus. In fact, it is an iterative process since goals often evolve over time. A personalized MHN system aims to capture, store, and keep track of the desired mental states and goals of the user to better inform the estimation and later the guidance procedures.    

The \textit{Monitoring} module in MHN is where a multimodal stream of objective and subjective information is collected from an individual to better understand his/her current mental state, context, lifestyle, adherence to guidance, etc. Building a navigation system is only possible by having an accurate estimate of current states based on measurements. In fact, what made route navigators successful was the integration of GPS which provides a fairly accurate estimate of current location.
MHN can leverage smartphones and wearable and portable sensors (e.g., smart rings, smart watches, smart speakers, etc.) that  map the response of bodily functions to mental health changes,
and identify normal bodily parameters to determine the mental health state in everyday settings.
The same technology also allows caregivers to monitor lifestyle (sleep, activity, context, life events, etc.) in order to identify and suggest modification of certain lifestyle events which can promote health. Some of these measurements are objectively measured (using sensors) while some are tracked using self-reports, for example, through smartphone apps. 
Since individuals interact continually with their smartphones, we can acquire the individual's  objective life log that reflects many aspects of everyday life. 
Furthermore, since people carry their phones all the time, it is possible to ask them to reflect on their life at anytime. 
This multi-dimensional stream of data can provide us with a holistic picture of an individual's status augmenting the traditional monitoring and therapy methods, for example by using a provider-in-the-loop model.

The next step in this cycle is the \textit{Mental Health Estimation} module which uses all the measurements collected by the Monitoring module to estimate the person’s mental health using quantitative as well as medical knowledge-based techniques. This estimation, for example, could indicate proximity to a mental disorder (e.g., anxiety depression, etc.). The objective of estimation would be to measure the mental health state without assigning it to a specific disorder. 

The \textit{Personal Models} module in MHN is intended to build and update models of physiological, psychological, and behavioral patterns for each individual to personalize the monitoring, estimation, and guidance provided by the system. 
In addition to the data generated by different modules in MHN (i.e., monitoring, estimation, and guidance), this module 
also needs to access 
a range of other personal data, including  prior mental/medical conditions experienced by the individual
and their genetics.
This module can grow in the number of components and dimensions to achieve more accurate and robust models as more information about the individual can be captured. 




The \textit{Guidance} module in MHN is in charge of interventions which depending on their types and risk-levels can be provided automatically using AI-based recommender systems (e.g., through smartphone or web application) or indirectly by a therapist in the loop. To change a person’s mental health state, the user (i.e., the individual or the healthcare provider) would issue a request and the system would use psychological, environmental, and other relevant knowledge sources to provide guidance in terms of lifestyle or environmental changes or ordering/modification of diet and medications for getting to the desired state. 
This information on previous guidance will be also integrated into the personal models.
However, providing guidance alone is not enough. Guidance must be situationally actionable and easy to follow. In most cases, mechanisms for nudging, incentivizing and inspiring might be required, along with subtle approaches for measuring compliance.

Mental health navigation must necessarily be personalized. While traditional psychological models are only focused on determining the differences between individuals, personalized psychological modeling has recently gained interest
\cite{insel2014,wright2020}, aimed at considering within-person changes as well. With these personalized models, we expect to have different solutions for the same mental state goal for different individuals. Personal models can also keep a profile of each person and improve that profile as they are using the system by tracking the person's performance given all historical guidance. 

\begin{figure} 
  \centering
  \includegraphics[width=0.8\linewidth]{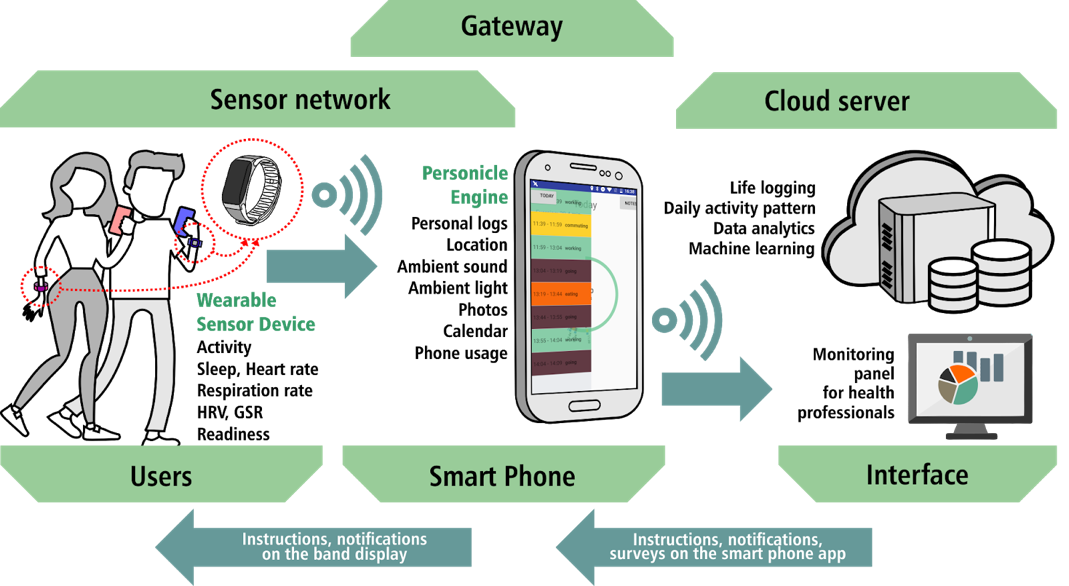}
  \caption{An exemplar 3-layer (sensor-gateway-cloud) architecture for implementing the MHN system}
  \vspace{-5mm}
  \label{fig:monitor-arch}
\end{figure}

To enable creation, updates and management of the personal models in everyday settings, a system implementing an efficient and layered architecture is required. Such system not only needs to monitor the individual, but also its context and lifelogs.   
The advent of ambulatory wearable devices that can unobtrusively capture and transmit physiological signals to servers when individuals are embedded in their own ecology opens the door for continuous capture of mental health responses, and offers a template architecture for implementing the personalized MHN system.
Figure \ref{fig:monitor-arch} depicts such an exemplar 3-layer sensor-gateway-cloud architecture comprised of:
i) a Body Area Network (BAN) with wearable devices, ii) an Internet-connected gateway (e.g., smartphones), and iii) a Cloud-enabled back-end support. 
At the Body Area Network (BAN) layer, a Wearable Internet of Things (WIoT-based) monitoring system leverages two main categories of sensors: 1) low-energy and inexpensive wearable sensors with continuous monitoring capability that measure signals such as photoplethysmogram (PPG), physical activities, heart rate (HR), heart rate variability (HRV), galvanic skin response (GSR), respiration rate (RR), and voice that enable physiological and behavioral assessment of mental health; and 2)
smartphone-based virtual and physical sensing features (for quantitative and qualitative capture of daily activities, life-logs, and context) that can further assist the system to understand and incorporate the subjects' personal lifestyles and behavioral patterns into their personalized healthcare.  We call such an engine, Personal Chronicle (i.e., \textit{Personicle} \cite{oh2017multimedia}) which is implemented in the second component.

The second component - the Internet-connected Gateway
- functions as the bridge between the BAN and the cloud. The smartphone implements two key services: the Personicle engine, and a foreground user interface. The Personicle engine (first service)
performs activity and context detection, signal preprocessing (e.g., filtering), data aggregation, and local storage. This engine is a passive background application that collects data from smartphone sensors (e.g., GPS, accelerometer, screen activity), executes tailored smartphone applications (e.g., communication logs, calendar), and interfaces to the connected wireless wearable sensors. The second service is a foreground user interface that performs two-way communication with the subjects and the cloud/caregivers. 

Finally, the third component -- the Cloud -- handles
computationally intensive and high-level tasks such as modeling, event mining, and statistical analyses, as well as the feedback and recommendation systems, while providing back-end services such as storage, control panels, and user interfaces for the end-users. In the following sections, we discuss each module in the MHN system in more detail. 


\section{Goals and Desired Mental State}

At the core of the MHN system are goals and desired mental states, as they may interact with any part of the MHN system (as seen in Figure \ref{fig:mhn}). Broadly speaking, mental states refer to the cognitions (thoughts, beliefs) and emotions (feelings, moods) that can comprise an individual's experiences -- these mental states can be reflected in an individual's behavior, such as in the choices they make or the way they treat others. Individuals are motivated by conscious and unconscious goals that are then reflected in changes within their physical and mental states \cite{maslow1958dynamic}. For example, when an individual feels cold, they may experience a physiological response such that it gives rise to unpleasant feelings and sensations within their bodies; to return to homeostasis, an individual will enact a series of actions in order to reach a more ideal state. Whether they are conscious or unconscious of the exact cause of the situation (i.e., thinking one is cold and must alleviate this), the event elicits a response within the individual that then leads to behaviors based on how the individual appraises the event \cite{scherer1999appraisal}. More broadly, individuals are motivated to seek enjoyable, hedonic states (i.e., time with loved ones, engage in hobbies) or will engage in states that involve challenge or stress (i.e., a workout, studying and training) if they recognize that it will benefit their long-term goals. Desired mental states can consist of basic needs (i.e., sleep, sustenance) or more complex goals (i.e., running for government to enact change) that altogether may play a role in the individual's health and well-being. Furthermore, inability to achieve these goals can cause discordance between one's current and desired mental states. Through the MHN system, the same individual, however, can monitor their own physical and psychological profiles through available tools that allow them to better understand, reflect, and initiate intervention. Seeking intervention and guidance from a provider can also help the individual work through the current states in order to attain desired states. The guidance can be provided using provider-in-the-loop models, smart recommendation systems, or a combination of both.  

In many circumstances, an individual seeks mental health services to address a specific mental health or self-improvement need or goal (identified by the self or by another person in the individual's life). The individual may present this goal to the provider at the outset of therapy. This goal is often unspecific or ill-defined (e.g., “I want to feel better about myself or my life,” “I just want to be normal”), so the provider seeks to understand this need or goal better through an information gathering process. This process often includes understanding more about the location and timing the individual experiences the undesired or opposite state of the goal state (e.g., disliking self), how the opposite state affects the individual, and how the individual would know they were in the desired goal state. The provider may ask questions of the individual to better understand the individual’s worldview and experiences, because each individual’s experiences are idiosyncratic. Each answer the individual provides helps the provider better hone in on an understanding of how the individual views themselves and their experiences.

Through this process, the provider may either come to agree with the individual, that the individual’s goal makes sense and is achievable or desirable (referred to as goal consensus, which could occur in the case where the individual wants to “feel better about themselves and life”) or the provider may begin to view things differently than the individual and may start to diverge in their perspective of the goal of the therapy (and then they must collaborate to revise the goals) \cite{tryon2011goal,tryon2018meta}. For instance, whereas the individual may “just want to be normal,” the provider may start to begin to hold a goal that the individual comes to accept themselves for who they are and embrace the unique qualities that make them this person \cite{bernard2013introduction}. In this process of goal consensus and collaboration, the provider shares these diverging perspectives with the individual in order for the individual to begin to see that the provider holds a different goal (that the provider sees qualities in the individual that are good and worthy of acceptance) and that the provider thinks they should work together toward that goal \cite{tryon2011goal,tryon2018meta}. Thus, the provider is suggesting a revision of the goal. The way in which this collaboration and divergence progresses varies by individual – for instance, with some clients, it might be able to be relatively rapid, but with others it might need to be a slow, iterative process of getting them to be open to seeing another perspective  \cite{aderka2012sudden,shalom2018intraindividual,shalom2020meta}. 
As discussed in Section \ref{sec.est}, the mental health state of an individual is defined in a high-dimensional space capturing different aspects ranging from behavioral traits to biological markers. 
Forming such a space and being able to estimate health states in it are essential to define and track both the current state and the desired state of an individual. 
This requires a holistic and comprehensive data acquisition procedure (i.e., the Monitor module) which is presented in the following section.  

\section{Monitor}
\begin{figure} 
  \centering
  \includegraphics[width=1.0\linewidth]{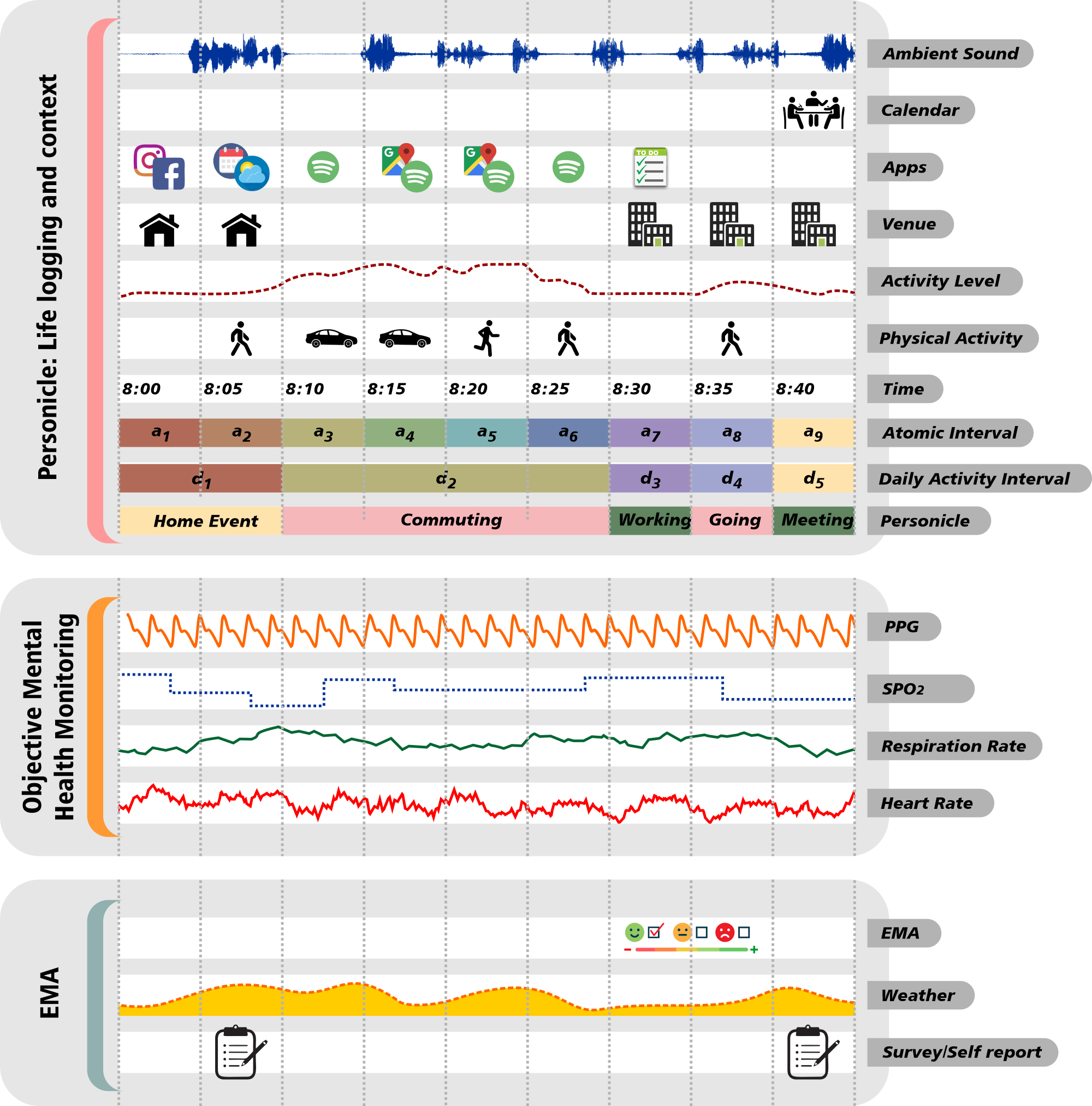}
  \caption{The main components of the Monitor Module including a potential multi-modal set of information sources}
  \vspace{-5mm}
  \label{fig:monitor}
\end{figure}

Monitoring is the main data acquisition module in the MHN model and serves as the gateway of information flow from the individual. This module can support different flows of information such as life-logs (Personicle), physiological signal assessment (wearable devices for objective measurements), and Ecological Momentary Assessments (EMA). 
Figure \ref{fig:monitor} shows an abstract representation of the Monitor's main components, including a potential set of multimodal data sources in the Monitor module. 
The following sub-sections discuss each component in detail.


\subsection{Personicle}
Nobel Laureate  Daniel Kahneman \cite{kahneman2004survey} postulated that quantifying a person's daily
usage of time into behavioral states (pleasure, stress level, etc.) could be beneficial for health research, resulting in several scientific approaches for monitoring and analyzing personal lifestyles and behavioral patterns. Table \ref{tb:activitylist} shows the list of Kahneman’s daily activities. Contemporary life event recognition technologies
include: i) situation specific recognition (e.g., smart home), ii)
computer vision-based recognition (e.g., surveillance or wearable
cameras), and iii) sensor-based recognition (e.g., accelerometer, and GPS). These enable capture of low-level lifelogs (e.g., step count, GPS, venue, or physical activity) but critically lack the ability to capture or infer important contextual and higher cognitive factors that enable assessment and prediction of a person's lifestyle. Our preliminary work on building a \emph{Personicle} (Personal Chronicle, shown at the top of Figure \ref{fig:monitor}) lays the foundation for a multi-modal life logging framework that integrates heterogeneous BAN sensory data (e.g., activity, location, phone oriented context, social interactions, etc.), synchronizes the data streams, recognizes some of Kahneman's daily activity, and builds a personal chronicle (\emph{Personicle}) of daily activity \cite{Jalali2013,jain2014,oh2015intelligent,oh2017multimedia,helal19}. 





\begin{wraptable}{r}{9.0cm}
\vspace{-0.5cm}
\small
\centering
\caption{List of Kahneman's daily activity}
\label{tb:activitylist}
\resizebox{.5\textwidth}{!}{%
\begin{tabular}{|c|l|c|c|}
\hline
\multicolumn{2}{|c|}{Still}                                                 & Working                                                          & Watching TV                                                          \\ \hline
\multicolumn{2}{|c|}{Walking}                                                        & Commuting                                                        & Preparing food                                                       \\ \hline
\multicolumn{2}{|c|}{Running}                                                        & Exercising                                                       & Socializing                                                              \\ \hline
\multicolumn{2}{|c|}{Cycling}                                                        & Religious event                                                  & Housework                                                            \\ \hline
\multicolumn{2}{|c|}{Driving}                                                        & Shopping                                                         & Intimate relations                                                   \\ \hline
\multicolumn{2}{|c|}{\begin{tabular}[c]{@{}c@{}}Direct communication\end{tabular}} & \begin{tabular}[c]{@{}c@{}}Eating\end{tabular} & \begin{tabular}[c]{@{}c@{}}Relaxing\end{tabular} \\ \hline
\multicolumn{2}{|c|}{\begin{tabular}[c]{@{}c@{}}Remote communication\end{tabular}}                                              & Using toilet                                                     &  Taking a break                                                            \\ \hline
\multicolumn{2}{|c|}{On the smartphone}                                                               & Home event                                                       &      Sleeping                                               \\ \hline
\end{tabular}%
}
\end{wraptable}


The Personicle component in the Monitor module 
objectively and automatically measures one's life events to help the MHN system incrementally model the person based on 
multi-modal life logging measurements.
The
\emph{Personicle} model itself must evolve using learning techniques
applied to daily activities in the chronicle and relate them to
biomedical or behavioral signals.
The \emph{Personicle} component enriches the other subjective and objective mental health related data of an individual with multi-modal contexts
around them to help build personalized models. At the lowest semantic
level of the \emph{Personicle}, the focus is on recognition of daily activity
(Figure \ref{fig:monitor}), as suggested by Kahenman et al. \cite{kahneman2004survey}. 
Even at this level, a major technical challenge is that fully-automated tracking does not
always  guarantee high recognition accuracy \cite{choe2017semi}. 


There exist a variety of approaches using different modalities to perform life-logging and activity recognition such as sensor-based, situation-specific, and computer vision based approaches (briefly reviewed in \ref{sec.ar}). 
We posit that a holistic approach leveraging multi-modal data together with these approaches can offer reliable recognition of the majority of daily activities listed in Table \ref{tb:activitylist}. In \ref{sec.ex_per}, we demonstrate an example of detecting an eating activity to discuss how Personicle exploits such a multi-modal approach to automatically detect certain complex daily activities. 

\subsection{Objective Mental Health Assessment}

\noindent
\textbf{Using Physiological Markers:} Ample research links activations of physiological reactivity with mental
health outcomes. For instance, elevations in sympathetic nervous system (SNS) activation, such as
increases in heart rate (HR), pre-ejection period, or electrodermal
activity (EDA) have all been linked with mental health outcomes 
\cite{borelli2018children,zahn89}. Moreover, greater decreases in
parasympathetic nervous system (PNS) activation (heart rate variability
{[}HRV{]}) have also been meaningfully associated with anxiety and
depression \cite{borelli2014school,hughes2000}. 
Similarly, objective bio-signal analysis to detect stress has been shown to be expandable to detect other psychological disorders, such as 
bipolar disorder that can significantly affect aforementioned physiological measurements \cite{valenza2014characterization,vishwanath2020investigation,vishwanath2020classification}.

Objective mental health assessment poses a challenge for
measurements  collected  during  standardized laboratory tasks that have high internal validity, but sacrifice considerable external validity in terms of generalizability to the types of experiences individuals encounter in their daily lives.  
The advent of ambulatory wearable devices that can unobtrusively capture and transmit to servers physiological signals when individuals are embedded in their own ecology represent significant advances in our ability to explore the stress response. 
In \ref{sec.stress}, we use objective stress assessment as an example that becomes  feasible 
using low-cost wearable sensors. 
In fact, there is a growing list of  consumer products (e.g., Samsung Galaxy watch, Fitbit Sense, Garmin vivosmart, etc.) that have already integrated different implementations of this service in their wearables.




\vspace{6pt}
\noindent
\textbf{Using Behavioral Markers:} Changes in our mental states also manifest themselves in our behaviors. Many of these behavioral changes can be automatically captured using technology. For instance, the fields of emotion recognition and affective computing boomed in 2010s thanks to advances in techniques such as facial expression recognition, audio-visual emotion recognition, and text mining and natural language processing for sentiment analysis. 
A summary of recent advances in multimodal affect recognition can be found in a report by D'Mello \textit{et al.} \cite{AFFECT} which includes applications such as automatic assessment of depression and anxiety. 
In regard to sensing, this field also overlaps with the field of activity detection and life-logging presented in the previous section.

Smartphones are rich sources of information for detecting life activities,
and can also partially reveal the mental state of an individual.
For instance, they can capture whether individuals are socially isolated by tracking their interactions (e.g., location, calls, social networks, etc.), or whether they are anxious or stressed out by analyzing their keyboard typing behaviors \cite{keyboard}. Smartphones have also shown to be effective in detecting certain mental disorders directly. 
For instance sleep distribution and quality can be detected by smartphones using screen usage, body movement, coughing and snoring, and ambient noise \cite{personalsensing}. 
Sleep qualities have been shown to be a useful indicator for different disorders such as depression \cite{studentlife} and chronic stress \cite{stresssleep}. 
Moreover, frequent mobile phone usage (on screen) has been correlated to  current symptoms of depression especially within young adult men and women \cite{phoneusage}. 
Further analysis has shown that frequent usage of mobile phone  could be a marker of early-stage sleep disturbances in  men, and depression in both  men and women. \cite{phoneusage}
Sustained and intense social network usage (high frequency and duration) on mobile phones
has shown to be an indicator of stress disorder and anxiety \cite{mobileusage,socialanx, socialandx2}. 
Phone addiction is also highly correlated with more specific behaviors such as impulsivity \cite{impulsivity} and social anxiety \cite{socialandx2,socialanx}.

Mobile device usage as a medium of interacting with the outside world is another rich source of information about a person's mental wellbeing.
The frequency, amount and recipients of an individual's phone calls and messages are a direct indicator of the individual's level of sociability and isolation. 
Furthermore, the content produced by an individual  can be processed to identify the individual's emotion throughout time. 
Several approaches have been proposed to recognize user emotions from textual content. These approaches are often  lexicon-based, machine learning-based, or hybrid \cite{shaheen2014emotion,sagha2017stacked}. 
Mobile devices have also been used
to detect certain disorders such as ADHD,  leveraging the phone's accelerometer and GPS outputs to understand the individual's movement in an environment to track ADHD footprints \cite{torous2016new}.




Since the physiological and behavioral modalities are complementary in nature, a holistic and multi-modal approach capable of properly fusing them can lead to  enhanced objective mental health assessments.


\subsection{Ecological Momentary Assessments (EMA)}
Although there are numerous technological advances in wearable sensors for physiological assessments in real-time and daily life, researchers have also considered advances in subjective reporting methods. Current mood and subjective states can influence recalled mood and states \cite{bradburn1987answering,tourangeau1999remembering}. One-time or global self-reporting of experiences may be susceptible to these biases and inaccuracy, as individuals might think of their overall or peak experiences \cite{bolger2003diary}. Thus, researchers have developed methods such as ecological momentary assessments (EMA) as a range of data collection techniques to capture in-the-moment experiences as they occur; EMA studies can reduce recall bias and allow for greater ecological validity, allowing individuals to report based on lived experiences as opposed to lab-based paradigms that may limit context and less accurately reflect real-world settings \cite{shiffman2008ecological}. Within the area of EMA, there are different ways in which daily assessments can help researchers understand people's lived experiences and behaviors, each with their different advantages and limits.

EMA as an umbrella term includes ambulatory assessments such as physiology and sleep, as well as frequent, multiple self-reporting of current states (i.e., experiencing sampling, daily diary). Some methods will prompt individuals at random times during the day to report their experiences, whereas others will either ask individuals to report at the end of the day or continuously monitor an individual's behavior using non-invasive wearable devices  (for a review \cite{bolger2003diary,shiffman2008ecological}). At their core, these methodologies collect multiple assessments for extended periods of time. Furthermore, the dynamics that occur within an individual's life point towards the importance of understanding how variation in an individual's experiences and states may relate to their mental health. Indeed, this examination of within-person dynamics reflect an individualized-approach where there is more variation within the person that is important to a nuanced understanding of behavior and mental states rather than group-level aggregates \cite{fisher2018lack,morin2011general}. These offer unique ways to understand the dynamics in which people's lives shift across time and the events that occur in their life. For example, EMA methods may be particularly relevant to understanding mental states and experiences during large-scale shared events such as a global pandemic, a natural disaster, an upcoming election, or, alternatively, more personal stressful life events such as soon-to-be parents, a licensing exam, or grieving over the loss of a loved one. Alternatively, researchers may consider how experiences such as drug-use or instances of relational conflict in life relate to their well-being. Overall, EMA methodologies can be effective, ecologically valid tools for researchers to use in understanding human phenomena.

A person-centered (i.e., within-person) approach essentially calls for more person-specific and individualized approaches to not only understanding human behavior but also psychopathology and development of interventions \cite{ebner2009ecological,wenze2010use,wright2020personalized} . Often, assessments of mental health symptoms require individuals to report in the past week or month. With daily diary and EMA methods, researchers can understand the experience of day-to-day events and mental states that co-occur with mental health symptoms, including depressive symptoms and their severity. EMA has been widely used in clinical research, including depression and mood disorders \cite{connolly2017rumination}, affective and emotion-related processes \cite{armey2015ecological,kuppens2010feelings}, and the intersection of psychopathology and emotion \cite{palop2010quantifying,silk2011daily,trull2008affective}. However, a few limitations to an EMA approach using self-report is that random prompting throughout the day may be inconvenient for individuals, and requires a heavy burden on individuals; daily diary methods that prompt individuals to complete at the end-of-day may also be susceptible to a degree of recall bias \cite{bolger2003diary}. Despite these limitations, much can be learned through an individualized approach to psychopathology and well-being. 

\section{Mental Health Estimation}\label{sec.est}
\begin{figure}[tb]
  \centering
  \includegraphics[width=0.35\linewidth]{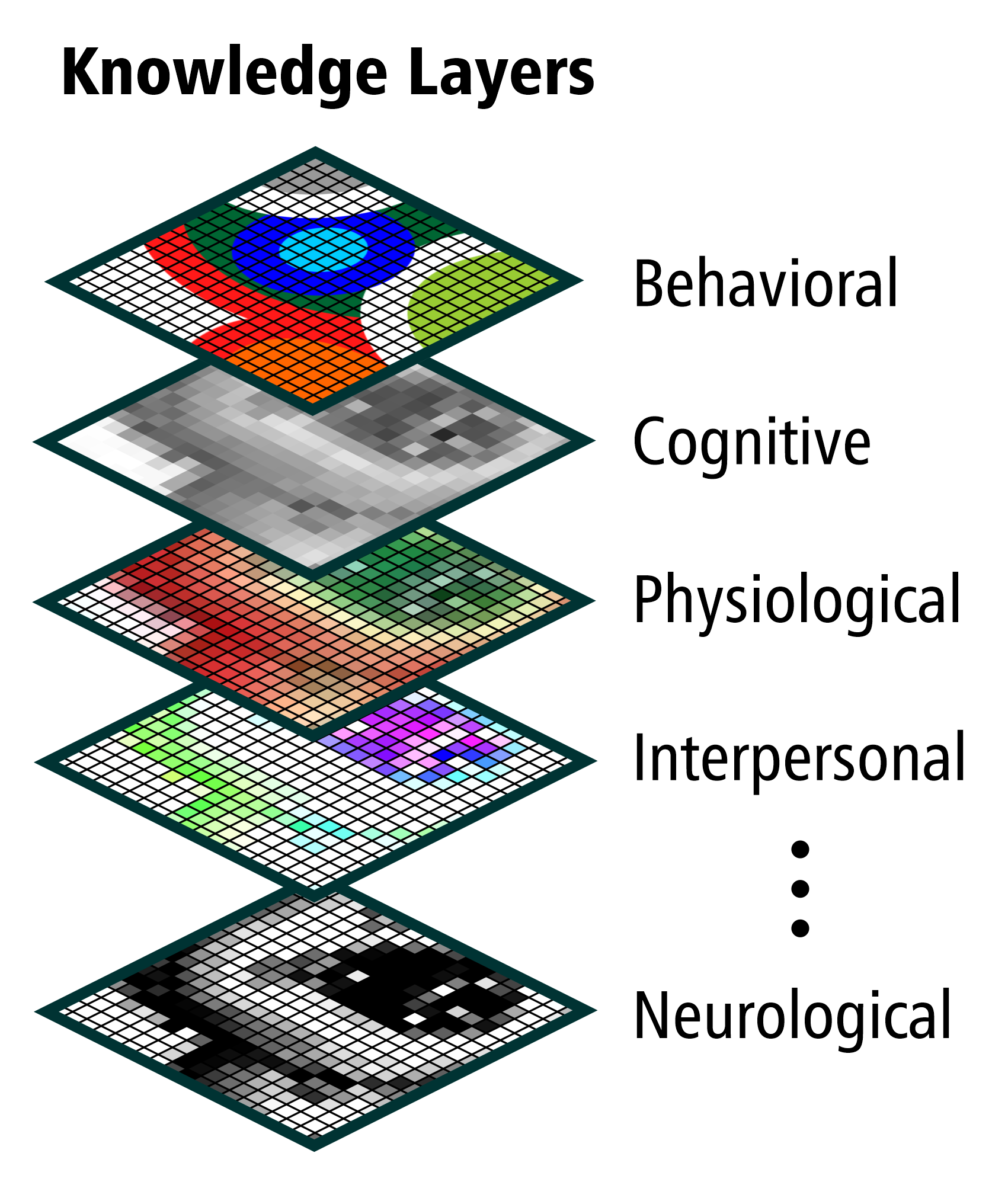}
  \caption{The knowledge layers in Personalized Mental Health Navigators}
  \label{fig:3D}
\end{figure}

The high-dimensional and holistic information collected by the Monitor module can be used to identify an individual's mental health state. Each individual is unique with large possibilities of mental health states. Many different dimensions need to be considered to model and estimate such a state, for instance, emotion, biological patterns, sociology, behavior, and cognition. Mental health state can be defined as a projection of covers across each dimension of the knowledge layers (Figure \ref{fig:3D}).
While a large body of research has been conducted to investigate each of these dimensions \cite{liang2019survey} (often in isolation), to the best of our knowledge, this field lacks a concrete method to define a general model of an individual's mental health state in a holistic way. Note that our daily life alters our state in such a high-dimensional space, simultaneously in many dimensions which can help determine our current state in the mental health continuum. 

Due to the complex nature of human psychology, mental health is a high-dimensional space, making the state estimating task challenging. 
One way to view it would be similar to a multi-layer coordinate system to describe the globe that leverages latitude, longitude, and altitude independently of any knowledge layers. 
In the context of mental health estimation, some relevant domain knowledge layers can be Emotional Factors, Behavioral Trait, Social Factors, Cultural Factors, Linguistic Factors, Cognitive state, Biological Markers \cite{liang2019survey}, Interpersonal relations, and, Neurological state (Figure \ref{fig:3D}). Almost any known psychological disorder can be identified by a combination of these dimensions. For instance, bipolar disorder, characterized by dysregulation of emotions \cite{wegbreit2015facial}, or depression \cite{ge2017social}, eating disorder \cite{levine2012loneliness}, sleep problems \cite{smagula2016risk} can be highly affected by many of such factors at the same time (e.g., social, neurological, cultural, etc.). 

A key functionality of the MHN system is to  process different modalities in order to properly estimate mental health variables in different dimensions. As mentioned earlier, this system has an actionable design given the individual's current state and goal state. It reflects on the desired mental states at each stage as mental health could potentially be viewed as a mismatch in one's current mental state and their desired mental states; and it is through the estimation, matching, and guidance that individuals come to an awareness of this mismatch which can result in help-seeking or self/other intervention.

\begin{figure}[tb]
  \centering
  \includegraphics[width=0.4\linewidth]{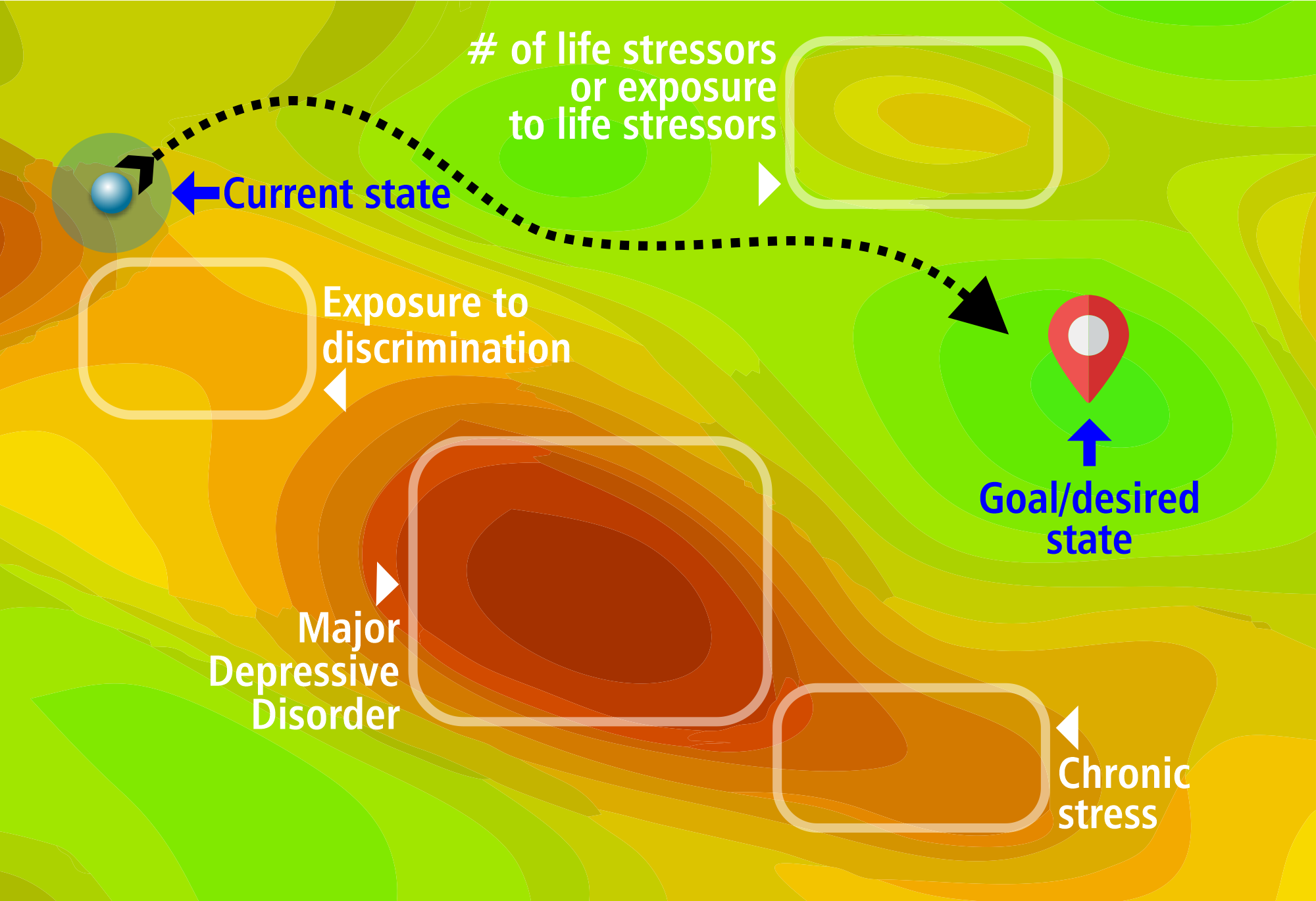}
  \caption{A sample mental health state w.r.t. emotional factors}
  \label{fig:2D}
\end{figure}


For providing guidance, it is important to know the current mental state and all adjacent and reachable states from this state.  Figure \ref{fig:2D} shows a much-simplified conceptual state-space in two dimensions.  It must be mentioned that the mental health state-space will be a very high dimensional space possibly requiring tens of different dimensions.
A person’s mental state at a time will be represented by a point in this space as shown in the diagram.  The state-space itself may be segmented to represent hyper-regions corresponding to different mental situations including mental disorders. As indicated in the diagram, different conditions like chronic stress and depressive disorders may be identified as a region in this space. 
As may be clear from this state-space, from the given current state, one could identify the path for navigation depending on the goal (ideal) state. 
Thus, mental health navigation is the problem of finding the most effective and realizable routes from the current state to the goal/desired state considering different constraints in this space.


\section{Personal Models}
Given that each person is unique, the MHN system leverages \emph{personalization} to incorporate the individual's unique genetics, lifestyle and exposure. Every individual is a biological system that responds differently to different inputs. This fact is the basis of the P4 approach to healthcare.  To build an approach that makes P4 a persistent action to guide the person in her/his lifestyle and medication, it is important to estimate and build its personal model.  
The  personal model is the heart and soul of the MHN system.
Since each person is unique, their personal model is essential to predict what is likely to happen to this person's mental health as a result of changing contexts and inputs.
Such predictions are not possible using population models.
If the prediction for the person suggests undesirable future outcomes, it is essential to prevent those by guiding the person to take right actions at the right time.  
Such guidance also requires using personal models. 
In other words, the personal model is central to all processes in the MHN system, from monitoring to guidance (Figure \ref{fig:mhn}). 
This necessitates building, updating, and reflecting on personal models at each and every stage of the cybernetics loop. 
Furthermore, the participation of the individual is essential for personal model creation, implementation, and guidance,
since 
which approaches may work on an individual and which may be ignored are again dependent on the person.


\vspace{6pt}
\noindent
\textbf{Building Personal Mental Health Models:} In many areas relevant to personalization, a model is built by collecting data in the context of the application.  Google, Facebook, Amazon, and Netflix are used as prime example of personalized recommendations using models built for each individual when they interact with different applications on their site.  
In the same way, most web applications increasingly log each and every interaction a user has on their site and the context of that interaction.  
These logs are then analyzed to build models of users that could predict and recommend to them what will be most relevant to them. 
Logs collected about user behavior on the site are the first step in data that is then analyzed to build models.  
This process of data logging and then analysis to build models is the basis of many scientific modeling.  
In most area of the sciences, the first step is to collect data in the context of analyzing a specific phenomenon for building a model that could be used for prediction and then control. MHN follows the same approach in mental health.

For building personal mental health models, one must log all relevant data related to mental state of a person. In case of MHN, a wide variety of data (mentioned in the Monitor section) is collected for an individual.  Such longitudinal data of an individual  then needs to be used to understand relationships among all events related to mental health.  In addition to the continuous data collection, a set of historical and demographic information of a patient's mental and medical conditions is gathered through questionnaires and other medical documents. These information could be anything from medical and psychological situations and surgeries to genetic information and history of diseases and disorders in the family.  These collected information are being used as context to gain a more accurate mental state of the patient. A Personicle is the log of everything that happens to a person and is collected over a long period.  This log is the major source of data that is analyzed using emerging event mining techniques for building both correlational as well as causal models for the person.

For instance, there are a set of mental disorders that are shown to have genetic roots. These could be found through OMICS analysis or by looking for the traces in the patient's family. Another example would be the effect of a heart condition on objective measurement of the stress that we expect more heartbeat regulations after a surgery. 

The personalized mental health modeling would also affect the desired mental state. Take the previous instance again: a patient who has undergone  heart surgery is generally expected to be more vulnerable to stress. As another example a patient with a history of bipolar disorder in the family statistically is expected to be prone to it, and this fact could affect the desired mental states.

Other examples where personal models can help would be learning and capturing the proper time and situation to trigger EMAs and self-reports or to send recommendations to individuals. Such models can help minimize the amount of missing data as well as increase the adherence to interventions.

\section{Guidance}
In this section, we elaborate on the \textit{Guidance} module of the MHN system. The \textit{Guidance} module represents the stage in which individual users of the MHN can synthesize and piece together information from the monitoring stage alongside their goals and mental states. More specifically, individuals can \textit{reflect} on their goals and progress, consider adjustments to their own behaviors or seek the \textit{recommendations} of the AI-based recommender systems or a healthcare provider. In summary, the individual at this stage can opt to \textit{intervene} in order to maintain current goal pursuit, form new goals, or alternatively revise goals in order to attain their desired mental states. Through the process of subjective and objective monitoring of one's health, individuals can review their current mental states and compare it to their desired mental states. When individuals are able to view their own data, they may have the opportunity to reflect on aspects of their daily life in which situations or feelings are less ideal. Furthermore, individuals can increase their own self-efficacy by taking an active role in their well-being and engage in self-management behaviors to maintain their health \cite{lorig2003self,villaggi2015self}. The empowerment and active role of the individual in their own management of health has been a shifting trend in the field of healthcare \cite{menichetti2016giving}. Indeed, the concept of self-management or self-monitoring has expanded into the field of digital mental health research, such that researchers have provided supporting evidence for its benefits in increasing emotional self-awareness and well-being \cite{bakker2018engagement}. Despite the emergence of digital spaces for increasing individual self-efficacy in engagement of health behaviors, self-monitoring and intervention in the trajectory of one's own mental health may still benefit from guidance and recommendations from others. 

Prolific amounts of state-of-the-art self-help and health-tracking apps are publicly available, yet limitations to many of these available tools is the lack of personalization (i.e., building personal models), guidance, or recommendations offered to the individual user of these apps \cite{caldeira2017mobile,epstein2015lived,ptakauskaite2018knowing}. 
Furthermore, the degree to which individuals engage and are capable of incorporating self-management strategies into their own lives for mental well-being varies significantly \cite{coulombe2016profiles}. A systematic review of health-apps available on the market indicated that goal-setting was crucial yet not always often available in these apps, particularly ones catered towards physical activity \cite{mckay2019using}. In particular, guided self-help may be beneficial for depression and anxiety \cite{moberg2019guided}. Additionally, Kelley and colleagues (2017) had conducted a two-part study to understand how individuals felt about self-tracking, where there may be harm if an individual became overly focused on their health tracking (i.e., caloric intake) and that initial scaffolding and recommendations during initial stages of tracking or monitoring may be necessary. Providers may initially struggle with interpreting the monitored data and incorporating it into useful recommendations \cite{ng2019provider}. However, with continued advances in self-monitoring and interest in these tools, providers and researchers are figuring out ways in which the patient-provider relationship can collaboratively utilize self-monitored information; indeed, providers may view the information as supporting self-motivation, better insight into the individual experience, opportunity for more personalized treatment plans, facilitating communication during the appointment \cite{chung2016boundary}. Thus, guidance may involve an important and dynamic process involving the self with one's healthcare provider. 

In collaboration with one's provider, the MHN system can improve the communication between individuals and their provider. Together, they can review the monitoring data and discuss changes in well-being along the mental health continuum. Based on the level of detailed information that includes subjective experiences, contextual and ecological information about the individual's living environment, as well as physiological fluctuations within the individual, a provider can offer professional recommendations or assist the individual in navigating and processing this information. The provider can then support the individual in developing self-efficacy in the engagement of health behavior changes, and prescribe treatment when necessary. Once the individual decides on a plan of action or intervention either with the AI-recommender system or from the help of a provider, the Guidance module then initiates the feedback loop where individuals can \textit{monitor} their experiences in the process of treatment and changes in behavior. Individuals again can consider their goals and mental states, and reconsider with their provider if the previously suggested recommendations or interventions were effective. At the return to this point in the system, individuals alongside their provide can consider alternative recommendations or interventions. 

\subsection{Utilization of Personalized MHN in Guidance and Treatment}

Consider the case in which a provider has access to data from the MHN system at their fingertips. Access to such data lends insight into the individual’s sleep, physical activity, physiology, stress levels, and activities the individual had engaged in throughout the week, as well as the interrelations between these variables. If the provider were able to integrate an understanding of these data into their conceptualization of the individual, how much further would the treatment of the individual progress and at what speed? This information could be used to augment and clarify the reports provided by the individual regarding their overall sense of the problem they are experiencing. In turn, this information could guide the provider’s recommended intervention to be more effective and tailored to the individual.

We offer two scenarios of two adolescent individuals who present at their psychologists’ offices with major depressive disorders – both are 17 year-old females who have been experiencing the following symptoms for the past four weeks: anhedonia (loss of interest or pleasure in things they used to enjoy), difficulty concentrating, feelings of guilt, suicidal ideation, difficulty falling and staying asleep, irritability, loss of appetite, and a sense of worthlessness. When both of these individuals complete a standardized clinical assessment of depression (e.g., the Beck Depression Inventory; \cite{beck1996beck}), they obtain the same score. However, if provided with the stream of data generated by Personicle and the MHN system, the psychologists working with these adolescents might be able to ascertain that these two individuals have very different profiles.

Individual A exhibits a profile characterized by high levels of social withdrawal – she has extremely limited contact with her family members, leaving her bedroom for meals and to use the restroom only, and spends the majority of her time in isolation or engaged in web browsing. She also does not appear to be engaging with her friends – she has not had face-to-face encounters or video-chat sessions with them nor has she texted them. Further, she experiences heightened physiological arousal during and after time spent on certain social media sites (e.g., Tik Tok, Instagram), which she most often does until late in the evening right up until the onset of sleep. Individual A may benefit most from an intervention approach that helps her reengage and make use of available social supports (family, friends) rather than watching from the sidelines through social media use. It is possible that there are conflicts in her relationships with her family or friends that Individual A does not know how to navigate – indeed, depressed adolescents often exhibit lower interpersonal skills  \cite{cruwys2014depression,nilsen2013social}. Individual A’s therapy could focus on acquiring interpersonal skills that would enable her to pursue these goals of reducing social withdrawal, which presumably would help her increase her mood and reduce her depression.

Individual B exhibits a profile that is characterized by high levels of relational turbulence. Individual B’s parents are divorced and she spends the weekdays at her mother’s home and the weekends at her father’s. Her mother and father are in the midst of a contentious legal battle and often send messages to one another through Individual B. Individual B cries four or five times a day, becoming hysterical at times and even hitting herself when she is extremely upset – during these fits of intense emotion, her heart rate and electrodermal activation become extremely high. She gets in yelling matches with both of her parents when they do things she thinks are unfair and has eloped from each of their homes once a week (for only an hour each time). She’s extremely protective of her younger brother and makes loom bracelets for him in her free time, exhibiting the lowest physiological arousal when she is in this state. She has found some friends on an online gaming platform to talk to while at her mother’s house– her parents forbid her from doing this, yet these friends are the only people she really feels she can connect with and talk to about her problems so she stays up late to talk to these friends when her mother is asleep. 

Individual B has a very different clinical presentation than Individual A, despite having the same symptoms of depression. Individual B may also have comorbid anxiety and/or emotion dysregulation. Thus, she may benefit first from some anxiety and stress reduction techniques, such as mindfulness-based stress reduction \cite{hofmann10} , progressive muscle relaxation \cite{holland1991randomized,jacobson1938progressive}, a treatment protocol that targets both anxiety and depression, such as the Unified Protocol for Emotional Disorders \cite{barlow2017unified,ehrenreich2009development}, or a treatment that addresses the relational context of her symptoms, such as Dialectical Behavior Therapy \cite{macpherson2013dialectical,linehan1993skills}. Alternatively, she may benefit from an attachment-based approach that involves her whole family system or part of her family, such as attachment-based family therapy \cite{diamond2002attachment}. A clinician would be looking at the types of therapies that emphasize emotion regulation and distress tolerance, particularly those that do so within an interpersonal context (because Individual A appears to become particularly dysregulated in interpersonal contexts). 

In reviewing these potential cases of depression manifested uniquely in two adolescents, we can observe the benefit in using individualized and personalized approach. Therefore, the MHN system and Personicle present an opportunity for mental health providers to make personalized treatment decisions that take into account these otherwise unaccounted for details.

\vspace{0.5cm}
\scalebox{0.8}{\begin{tabular}{|c|p{.42\textwidth}|p{0.42\textwidth}|}
\hline
    --& \textbf{Individual A}&\textbf{Individual B}\\ \hline

   \centering Contextual Information
    &
    \begin{itemize}[noitemsep,topsep=0pt,leftmargin=*]
        \setlength{\itemsep}{0pt}%
        \item High levels of social withdrawal - Limited contact with her family members
        \item Leaves bedroom for meals and to use the restroom only 
        \item Spends the majority of her time in isolation or engaged in web browsing. 
        \item No face-to-face encounters or video-chat sessions with friends 
        \item Experiences heightened physiological arousal during and after time spent on certain social media sites (e.g., tik tok, instagram), which she most often does until late in the evening right up until the onset of sleep. 
    \end{itemize}
    &
    \begin{itemize}[noitemsep,topsep=0pt,leftmargin=*]
        \item High levels of relational turbulence. 
        \item Individual B’s alternates between father’s and mother’s home 
        \item Mother and father in midst of contentious legal battle and send messages to one another through Individual B. 
        \item Individual B cries four or five times a day, hitting herself when she is extremely upset – during these fits of intense emotion, her heart rate and electrodermal activation become extremely high.
        \item Gets into yelling matches with both of her parents 
        \item Has eloped from each of their homes 
        \item Spends time making loom bracelets for her younger brother in her free time, exhibiting the lowest physiological arousal at this time. 
        \item Chats with friends on an online gaming platform while at her mother’s house– parents forbid her from doing this, but she has unlocked the site and chats after her mother goes to bed.

    \end{itemize}\\ \hline

    \centering Recommended Intervention
    &
    \begin{itemize}[noitemsep,topsep=0pt,leftmargin=*]
        \item Therapy focused on enhancing social skills
        \item Approach that helps her reengage and make use of available social support (family, friends) rather than watching from the sidelines through social media use.
    \end{itemize}
    &
    \begin{itemize}[noitemsep,topsep=0pt,leftmargin=*]
        \item Mindfulness-based Stress-reduction
        \item Progressive muscle relaxation
        \item Unified protocol for emotional disorders
        \item Dialectical Behavior Therapy
        \item Attachment-based Family Therapy
    \end{itemize}\\ \hline
    
\end{tabular}}

\section{Case Study} \label{case_study}

To examine the feasibility and possibility of understanding mental health through a personalized mental health navigation model, we conducted a pilot study. The purpose of this pilot study was to provide some preliminary data to examine the utility of this system and to help us envision future avenues into which to incorporate the MHN system. We recruited a small sample of young adults (college students) to complete baseline assessments of mental health and followed them continuously over the course of 12-months, conducting mental health assessments at regular intervals. Here we present a case study \cite{borelli2020nine} from our pilot data collection that reflects relevant aspects of the MHN model.

We focus on young adults because emerging adulthood (between the ages 18 and 25) is a stage characterized by significant changes \cite{hoyt2012positive,steger2009meaning} in routine and lifestyle as well as risk for the onset of many major health problems \cite{auerbach2018world,twenge2019age}. College students, in particular, are a subset of young adults \cite{mayhew2016college} who are exposed to a unique constellation of circumstances -- less parent supervision \cite{mulder2002leaving}, more peer influence \cite{mason2014social,patterson2019social,rinker2016social}, and greater sleep dysregulation \cite{becker2018sleep}. As a group, college students are less likely to seek support from mental health professionals \cite{eisenberg07} or professors \cite{eisenberg09}. Because this is a time when many individuals are also honing their skills for adaptive living \cite{lawrence17}, it is also an important time in which to intervene because habits can be formed that can last a lifetime. Furthermore as previously mentioned, young adults frequently use technology in their daily lives and could possibly benefit from smart devices to help monitor their health. 

\begin{figure}
  \centering
  \captionsetup[subfigure]{labelformat=empty}
  \addtocounter{subfigure}{-1}
   \begin{subfigure}{}
     \centering
     \includegraphics[width=\textwidth]{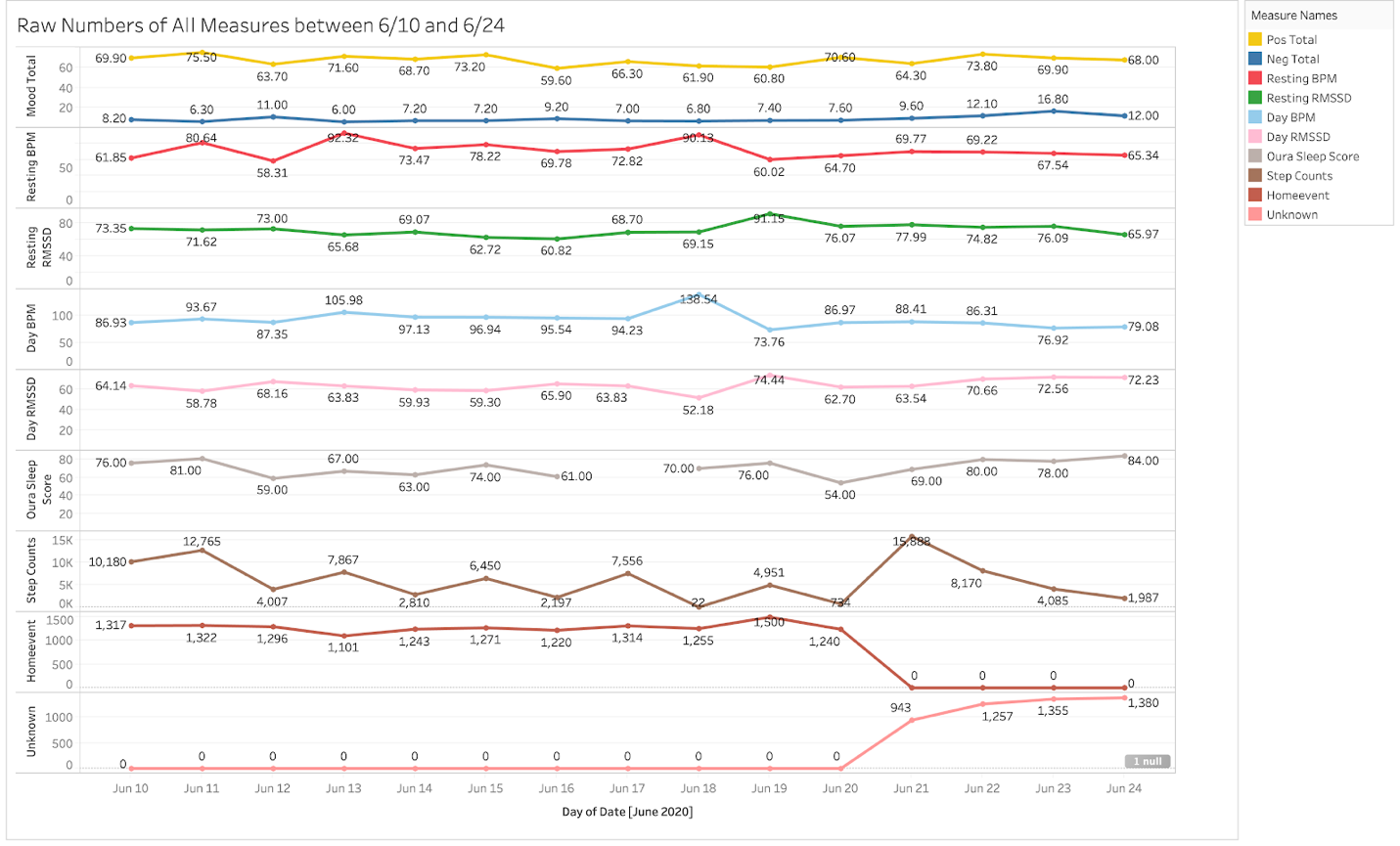}
     \caption{Well-Being (A)}
     \label{fig:case_well}
   \end{subfigure}
   \begin{subfigure}{}
     \centering
     \includegraphics[width=\textwidth]{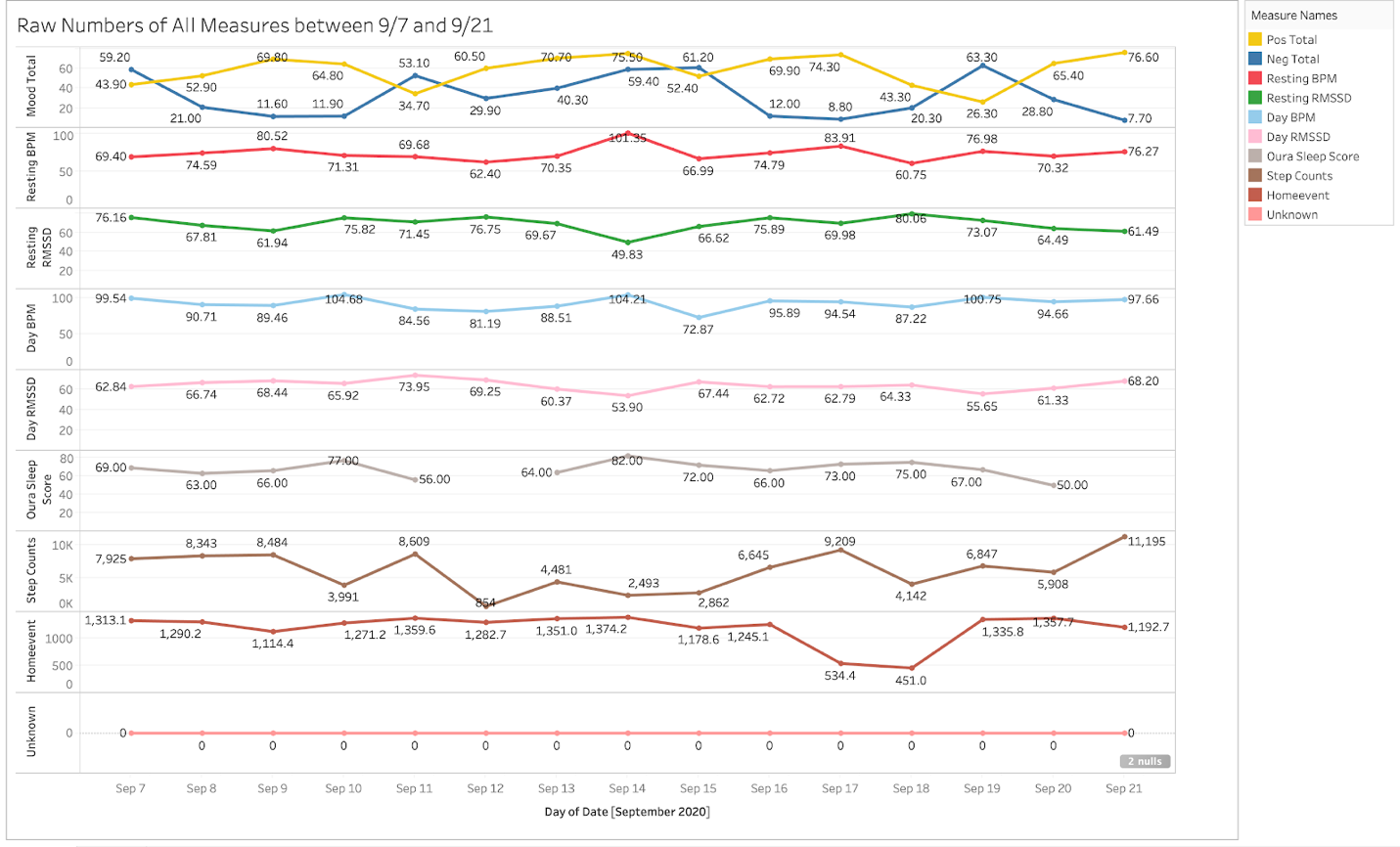}
     \caption{Poor Mood (B)}
     \label{fig:case_poor}
   \end{subfigure}
  \caption{Data Collected From Mental Health Navigator During a Period of Well-Being (A) and Poor Mood (B). Mood Total = Blue line signifies negative emotion and yellow line signifies positive emotion; Resting BPM = resting beats per minute, sampled 15 minutes every two hours, including sleep; Resting RMSSD = index of heart rate variability, sampled 15 minutes every two hours, including sleep; Day BPM = resting beats per minute, sampled 15 minutes every two hours, not including sleep; Day RMSSD = index of heart rate variability, sampled 15 minutes every two hours, not including sleep; Oura Sleep Score = summary score of overall sleep quality, a weighted sum of sleep contributors, which combines sleep latency, onset, restfulness, REM, deep sleep, timing and efficiency; Step Counts = generated from the watch, total number of steps taken per day; Home Event = minutes participant spent at home; Unknown; type of location of participant is unknown based on GPS}
  \label{fig:case}
\end{figure}

To illustrate the potential utility of the MHN model, we focus on one participant, a college-aged male, who entered our study with a depression score in the normal (non-clinical) range on the Beck Depression Inventory (\cite{beck1996beck} BDI-II; a score of 8) in January 2020, but by April of 2020, had a score of 24, indicating moderately severe depression. The COVID-19 pandemic hit the United States between these two time periods. 

Using the MHN model allowed us to not only \emph{monitor} his behaviors and states using a smart ring, smart watch, and phone app, but also collect his daily mood and a weekly open-ended text response regarding how he felt. Examining these aspects of \emph{monitoring} together follows the next component of the MHN model, \emph{estimation}. His narrative responses to our weekly survey questionnaire indicated that the COVID-19 pandemic had negatively impacted his life and well-being, as well as that of his family. Although his overall depressive symptoms increased dramatically over the course of the study, during the period of time when his symptoms were higher, there were phases when his mood states indicated that he was experiencing high mood (more positive emotion and less negative emotion) and other times when his mood was worse (when he was experiencing more negative emotion and less positive emotion). During the phase in which he was experiencing particularly poor mood (see Figure \ref{fig:case}), we observed a pattern wherein he was sleeping less, his heart rate variability was lower, he was less physically active, and he spent more time at home. However, during the good mood phase, he slept more, exhibited higher heart rate variability, he was more physically active, and he was outside of the house for greater periods of time. 

In the way, monitoring the participant through wearable sensors and collecting daily subjective experiences along with contextual information can help to the provider to understand factors that relate to the individual's overall well-being. The data and the individual’s mental states (i.e., how they are feeling) can help the provider consider what recommendations may be most beneficial. For example, potential recommendations for this participant based on their pattern of behaviors may include encouragement and maintenance of habits and interactions that seem to be best for their mood and well-being, such as physical activity and social interaction with friends. In their weekly feel-in reports, they reported feeling productive and purposeful when involved in work that benefited others. The provider can then reflect these back to the individual and help guide them to pursue lifestyle changes or maintenance in order to sustain a sense of purpose and enjoyment in their activities. In the future, we hope to develop the MHN to the point that it can be used to guide intervention as well -- in other words, the MHN could be deployed as a system in which the provider could introduce new inputs/suggestions into the individual's behavioral repertoire. The provider may wish to trigger in-app interventions to be delivered to the individual when the provider notices certain warning signs in the individual -- for instance, when the individual's positive and negative emotions show large discrepancies, this may be a time when the individual is in need of a reminder to engage in physical activity, reconnect with friends, or utilize an mhealth intervention available to them (e.g., an app that encourages them to savor memories of positive interpersonal connection).

Thus, our pilot data and the aforementioned case study provides evidence for the feasibility of the \emph{monitoring} and \emph{estimation} components of the MHN model, where we are able to collect objective and subjective data in real-time that captures the physical and mental states of an individual while also accessing this data to \emph{estimate} their well-being (i.e., shifts in their sleep and emotions). Based on the estimations and interpretations of their data, we offered a few recommendations that fall under the  \emph{guidance} aspect of the MHN model. \emph{Monitoring} an individual over the course of the past year allowed us to consider the therapist-in-the-loop (here, one of our authors who is a trained clinical psychologist) who reached out to the individual when their mental health data indicated an increase in their depressive symptoms to offer additional assessment and suggestions for intervention. In summary, our pilot data positions us well to capture the \emph{monitoring}, \emph{estimation} components, and to incorporating \emph{guidance} as part of an intervention in future studies.

\section{Conclusions and Future Work}\label{results-from-prior-nsf-support}
In this paper, we proposed the notion of Personalized Mental Health Navigation (MHN) as a goal-based cybernetics system allowing for a continuous cyclic loop of measurement, estimation, guidance, and influence to monitor and make sure that the person’s mental health state remains in a healthy zone. MHN has the potential to transform mental health care from reactive and episodic to a continuous and navigational paradigm that leverages a holistic, personalized model of the individual. As we move toward a world where we hope to be able to match each individual with a personally tailored treatment plan, finding ways to use technology such that it supports mental health providers in treatment planning and collaborative goal setting is a priority. The MHN offers one vision of what this could look like, wherein providers could have a round-the-clock inside view into their clients' lives, rendering it so that in-session interactions and recommendations can be supported by knowledge gleaned from out-of-session data points. There is work yet to be done -- we must chart the landscape for how to integrate these data sources so that they can be delivered to providers in ways that best inform treatment planning and collaborative goal-setting (this might differ somewhat by provider preference or therapeutic treatment modality, by the client's presenting problem, or by the client's behavioral patterns), identifying ways to distill the information received and provided to a manageable amount (for the provider and the individual receiving treatment), and determining how to take this to the next level so that providers can deliver interventions to their clients in real-time using the MHN. Each of these steps will require significant and careful investigation, but the promise of this approach cannot be overstated, as it offers the provider and the individual seeking treatment what they seek to give to and to elicit from one another, but always seemingly fall short of -- an unobtrusive bird's eye view into the client's functioning throughout their daily life. 

\bibliographystyle{unsrt}
\bibliography{draft}

\appendix
\section{Related Works on Activity Recognition Techniques}\label{sec.ar}
\noindent
\textbf{Sensor-based Activity Recognition:} Signals collected from an individual can be used directly for building current state model while they can also be used to understand the high-level context of his/her life.
Human activity recognition using wearable sensors and mobile phones has recently attracted much
attention yielding a rich body of research on activity recognition using
data-driven approach. More recently, activity
recognition has moved from deploying a computer-based toolkit to
exploiting mobile/wearable sensor devices. A comprehensive survey on
activity recognition using those devices \cite{incel2013review} shows that
current techniques mostly capture low-level physical motion of a user,
such as walking, running, and bicycling. Some research groups have
proposed rule-based reasoning with an ontology language so that they can
recognize more complex activities than physical activities \cite{riboni2011owl, riboni2011cosar, helaoui2013probabilistic}, such as showering, or meeting;
however, data used to recognize activities (i.e., conference room) is
difficult to collect in real-world scenarios.

\vspace{6pt}
\noindent
\textbf{Situation Specific Activity Recognition:} Research on
situation-specific capture has drawn much attention in Activity of Daily
Living (ADL) recognition. For instance, to automatically recognize ADL (e.g.,
toileting, bathing, showering or sleeping, etc.) researchers
have set up low-cost sensors at critical locations in a home \cite{chen2012knowledge, fogarty2006sensing, intille2002just, mittal2013versatile, tapia2004activity, van2008accurate} and then have
predicted activities using a Naive Bayes classifier \cite{luvstrek2015recognising, tapia2004activity}, ontologies
and semantic reasoning \cite{chen2012knowledge}, and Formal Concept Analysis
\cite{mittal2013versatile}, etc. 

\vspace{6pt}
\noindent
\textbf{Computer Vision based Activity Recognition:} P. Wang et al. \cite{wang2012semantics,wang2014enhancing,wang2016characterizing}
have highlighted the importance of visual lifelogs as they identify
various semantic concepts across individual subjects. They suggest
aiding
life logging solutions using cutting edge components (i.e.,
gathering, enriching, segmenting, keyframe, annotation and narrative)
for extracting meaningful knowledge from one's lifelog data. They
automatically identified high-level human activities such as eating,
drinking, or cooking using SenseCam images and data models. 


\vspace{6pt}


\section{Eating Activity: An Example of Using Personicle for Complex Activity Recognition}\label{sec.ex_per}

Eating events cannot be recognized in a straightforward way when a
user deviates from daily routine. Likewise, similar stop-and-go patterns can misclassify shopping or using the
toilet event. Personicle uses heart rate from wearables as a new attribute
with a hypothesis that the pattern of heart rate can be affected by
eating and using the toilet event. It also assumes that the pattern of
heart rate may have a different baseline on different timebands (e.g.,
morning, afternoon, evening), and thus handles them separately.
Lastly, it uses location attributes in order to restrict the range of
daily activities. For example, using the toilet, working, taking a
break, and relaxing can happen at the workplace. Eating, using the
toilet, housework, preparing food, and watching TV can happen at home.
It elaborates this on taxonomy as well as builds a daily activity
ontology in order to build a robust daily activity model. Once it figures out all the relationships or sequential orders among the daily activities through taxonomy and ontology, it can then easily recognize daily activities like “preparing food” if there is some amount of movement before eating at home. To do so, it considers the decrease of atomic-interval size from 5 minutes to 1 minute so that it can recognize more latent daily activities with detailed granularity. With the narrowed down atomic interval, and all the new measuring attributes, it finally re-formulates daily activity D by a triplet $T$ = ($D,A,R$), where $A$ is a set of attributes, and $R$ is the binary relationships between $D$ and $A$, $R$ $\subseteq$ $D$ $\times$ $A$. Once each daily activity is defined by the triplet, the triplet can then be converted into a cross table (e.g., Table \ref{tab:cross_table}). Then, all possible formal concepts ($X_i$, $Y_i$), where $X_i$ $\subseteq$ $D_i$, and $Y_i$ $\subseteq$ $A_i$, are extracted from the cross table, and then are set up as nodes in the concept lattice, which is a graphical representation of the partially ordered knowledge. The hierarchy of formal concepts can be constructed by the following relations: 
\begin{eqnarray} 
(X_{1}, Y_{1}) \leq (X_{2}, Y_{2}),\, if \, X_{1} \subseteq \ X_{2} \leftrightarrow Y_{1} \supseteq Y_{2} 
\end{eqnarray} $X_i$ and $Y_i$ satisfy the following relations: 
\begin{eqnarray} 
X_i^{'} &=& \{a_i \in A_i \mid \forall d_i \in X_i, (d_i,a_i) \in R_i \}\\
Y_i^{'} &=& \{d_i \in D_i \mid \forall a_i \in Y_i, (d_i,a_i) \in R_i \}
\end{eqnarray} 

It then builds classifiers for each daily activity by using the ensemble
technique, which has shown promising results in the previous research \cite{oh2017multimedia}.

\begin{table}
\small
\centering
\caption{Simplified cross table defining relationships between daily activity and their attributes.}
\label{tab:cross_table}
\resizebox{.5\textwidth}{!}{%
\begin{tabular}{l|l|c|c|c|}
\cline{3-5}
\multicolumn{2}{l|}{\multirow{2}{*}{}}  & \multicolumn{3}{c|}{\textbf{Attribute}}                                                                                                                                                                                                                 \\ \cline{3-5} 
\multicolumn{2}{l|}{}                                                          & \begin{tabular}[c]{@{}c@{}}Walking\\ (Experiential)\end{tabular} & \begin{tabular}[c]{@{}c@{}}Medium time-duration\\ (Temporal)\end{tabular} & \begin{tabular}[c]{@{}c@{}}Work\\ (Spatial)\end{tabular} \\ \hline
\multicolumn{1}{|l|}{\multirow{3}{*}{{\rotatebox{90}{\textbf{Object}}}}} & Working      &                                                                                  & X                                                                                         & X                                                                        \\ \cline{2-5} 
\multicolumn{1}{|l|}{}                                 & Using Toilet & X                                                                                &                                                                                           & X                                                                        \\ \cline{2-5} 
\multicolumn{1}{|l|}{}                                 & Commuting    & X                                                                                & X                                                                                         &                                                                          \\ \hline
\end{tabular}%
}
\vspace{-0.35cm}
\end{table}

\begin{figure}
  \centering
  \includegraphics[width=0.35\linewidth]{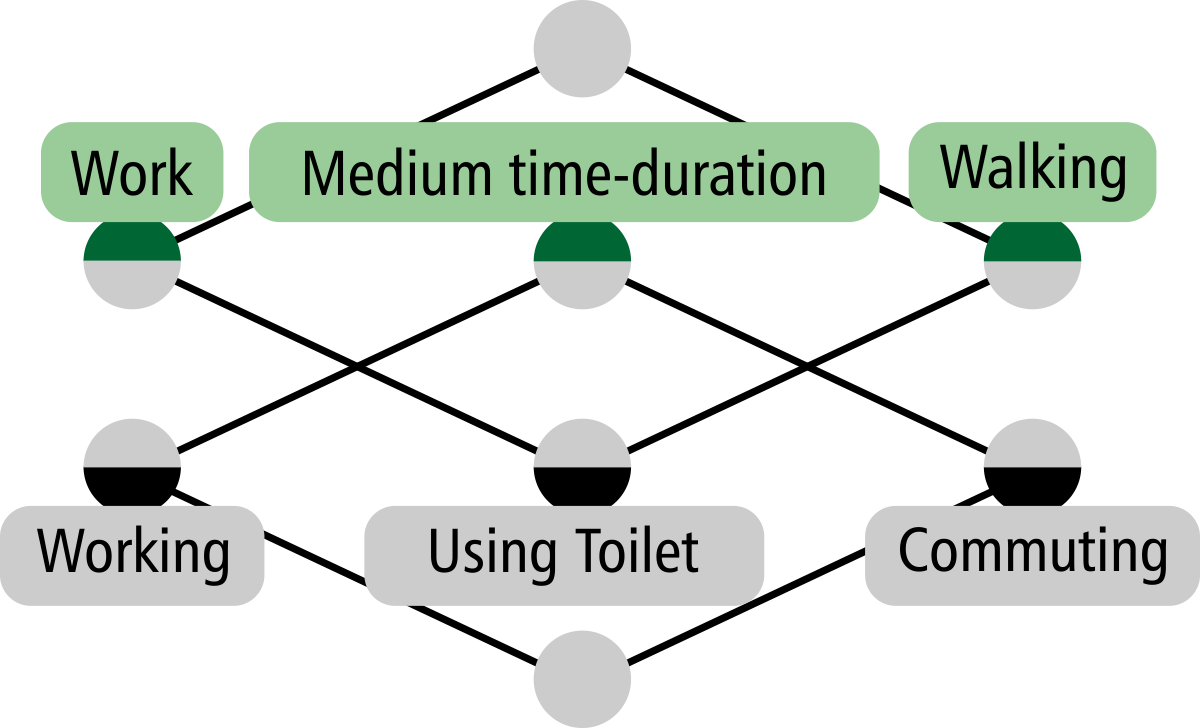}
  \caption{Sample concept lattice derived from Table \ref{tab:cross_table}.}
  \vspace{-5mm}
  \label{fig:cross}
\end{figure}


\section{Stress Assessment: An Example of Using Physiological Markers for Objective Mental Health Assessment}\label{sec.stress}
Stress takes three forms: 1) Acute or short-term stress, 2) Episodic acute (acute stress that
occurs more frequently and/or periodically), and 3) Chronic stress (caused by long-term stress factors). Stimulated by the SNS activation, acute stress causes a flight-or-fight response to stimuli and is not considered harmful in the short-term, but when its frequency increases, physiological consequences might occur. Episodic acute stress is associated with a busy and chaotic life and can be considered to be harmful when it occurs over prolonged periods of
time. 


Extracting cardiovascular measurements e.g., heart rate HR, resting heart rate (RHR), and
HRV from blood volume pulse (BVP) measured by PPG sensors as a low-cost, noninvasive optical technique can help detect stress level in an effective way (shown at the middle of Figure \ref{fig:monitor}). HRV variables change in response to stress. For
instance, lower HRV has been found to be associated with stress \cite{kim2018}. 
Time-domain variables of HRV show the
amount of variability in measurements of the interbeat interval (IBI),
which is the time period between successive heartbeats \cite{shaffer2017}.
In addition, the LF and HF components are
respectively associated with SNS and PNS
activities in the nervous system \cite{healey2005detecting, acharya2006heart}. 
The analysis involved in assessing frequency-domain HRV analysis
lies in the energy ratio of LF to HF content (i.e., the ratio of SNS to
PNS activity). The most frequently reported factor associated with
variation in HRV variables was low parasympathetic activity, which is
characterized by a decrease in the HF and an increase in the LF \cite{kim2018}.

\begin{figure}
  \centering
  \includegraphics[width=1.0\linewidth]{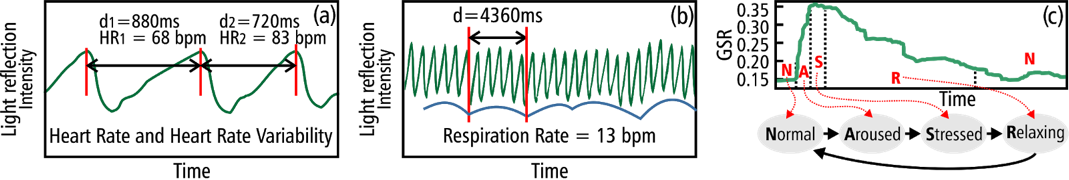}
  \caption{Stress-related physiological signs. (a) heart rate and HRV, (b) respiration rate, and (c) GSR.}
  \vspace{-5mm}
  \label{fig:stress}
\end{figure}

Monitoring respiratory rate (RR) and respiratory pattern as a predictor
chronic stress \cite{choi2010ambulatory}. RR is extractable from BVP in
two ways: using amplitude changes (the low-frequency component in BVP
shown in Figure \ref{fig:stress}(b)), and changes in the time between heartbeats.
Inhalation accelerates the heart rate, while exhaling decelerates it. In
addition to the stress-related insights extracted through digital signal
processing, researchers have also utilized machine learning to analyze and
classify stress level using the features discussed above \cite{sina19}.

Galvanic Skin Response (GSR) has also shown to be a rich physiological source of objective information. Through EDA analysis of the GSR signal, the inner process of stress can
be categorized into four states. It can be seen as a state of emergency
that is preceded by arousal due to an external stimulus, see Figure
\ref{fig:stress}(c). After the stressor disappears, the body relaxes and returns to a
normal state. This can be analyzed in different ways such as a
traditional classification task, as one-class classification, as event
identification, and as time series subsequence classification \cite{bakker2011s}. Borelli \cite{borelli2017school,borelli2015anxiety,panzariello} has also shown in an in-lab
setting that EDA can be a predictor of stress. However, several studies
have indicated that in everyday settings, several factors such as
person-to-person as well as day-to-day variations (e.g., due to ambient
temperature or limb movement) can result in low correlations in certain
situations showing the necessity for multi-modal and personalized objective stress assessment methods.

\end{document}